%% file: main.tex
\documentclass[10pt,conference]{IEEEtran}

\AtBeginDocument{%
  \providecommand\BibTeX{{%
    \normalfont B\kern-0.5em{\scshape i\kern-0.25em b}\kern-0.8em\TeX}}}

\usepackage{amsmath,amssymb,amsfonts}
\usepackage{algorithmic}
\usepackage{graphicx}
\usepackage{textcomp}
\usepackage{xcolor, colortbl}
\usepackage{booktabs}
\usepackage{multirow}
\usepackage{xspace}
\usepackage{caption}
\usepackage{subcaption}
\usepackage{footnote}
\usepackage{url}
\usepackage{tcolorbox}
\usepackage[para]{footmisc}
\usepackage{array,ragged2e}
\usepackage{multicol}
\usepackage{listings}
\usepackage{balance}
\usepackage{tipa}
\usepackage{rotating}
\usepackage{fontawesome5}
\usepackage{xurl}
\usepackage[hidelinks]{hyperref}

\newboolean{showcomments}
\setboolean{showcomments}{true}         
\ifthenelse{\boolean{showcomments}}
  {\newcommand{\nb}[2]{
  \fbox{\bfseries\sffamily\scriptsize#1}
     {\sf\small$\blacktriangleright$\textit{\textcolor{red}{#2}}$\blacktriangleleft$}
   }
  }
  {\newcommand{\nb}[2]{}
   
  }

\newcommand{\ie}{\emph{i.e.,}\xspace}
\newcommand{\eg}{\emph{e.g.,}\xspace}

\newcommand{\etal}{\emph{et~al.}\xspace}
\newcommand{\secref}[1]{Section~\ref{#1}\xspace}

\newcommand{\figref}[1]{Fig.~\ref{#1}\xspace}

\newcommand{\tabref}[1]{Table~\ref{#1}\xspace}

\newcommand{\commitsStudyOne}{914\xspace}
\newcommand{\relevantCommitsStudyOne}{230\xspace}
\newcommand{\assistantsStudyOne}{17\xspace}

\newcommand{\commitsStudyTwo}{1,487\xspace}
\newcommand{\prsStudyTwo}{760\xspace}
\newcommand{\issuesStudyTwo}{2,008\xspace}
\newcommand{\filesStudyTwo}{2,236\xspace}
\newcommand{\linksStudyTwo}{6,491\xspace}
\newcommand{\validConversationsStudyTwo}{299\xspace}
\newcommand{\validRoundsStudyTwo}{1,863\xspace}

\newcommand{\practitionersStudyThree}{20\xspace}

\newcommand{\categories}{87\xspace}

\definecolor{Gray}{gray}{0.9}

\def\BibTeX{{\rm B\kern-.05em{\sc i\kern-.025em b}\kern-.08em
    T\kern-.1667em\lower.7ex\hbox{E}\kern-.125emX}}
    
\ifCLASSOPTIONcompsoc
  \usepackage[nocompress]{cite}
\else
  \usepackage{cite}
\fi 

\begin{document}

\title{AI-Generated Code Issues: A Taxonomy from Repository Histories, AI Conversations, and Developer Interviews}


\IEEEtitleabstractindextext{%
\begin{abstract}
Large language models (LLMs) are increasingly used to generate source code, yet the quality issues in such code remain only partially understood. Understanding these issues can help developers adapt quality assurance practices, such as code review and testing, to better detect problems that may be more common in AI-generated than in human-written code. 
We study how quality issues in AI-generated code manifest in real software development activities through a mixed-method empirical study that draws on evidence from three complementary sources. First, we mine GitHub commits whose messages explicitly attribute a fix to a problem in AI-generated code, and we manually inspect the \commitsStudyOne matched commits to remove false positives and classify the quality issue being corrected. Second, we collect developers' shared conversations with AI tools, identify conversations including source code and textual signals of model mistakes, and manually analyze \validConversationsStudyTwo of them to classify the underlying quality issue. Third, we interview \practitionersStudyThree practitioners and researchers who use AI coding tools to enrich and validate the obtained catalog of quality issues. From these analyses, we derive a taxonomy of \categories types of developer-attributed quality issues in AI-generated code, going beyond functional bugs. 
Our findings contribute an empirically grounded taxonomy and implications for the design and evaluation of AI coding assistants and agents.
\end{abstract}

}

\author{
\IEEEauthorblockN{
Rosalia Tufano\textsuperscript{\textdagger}, 
Nargiz Humbatova\textsuperscript{\textdagger}, 
Daniele Bifolco\textsuperscript{\textdaggerdbl}, 
Gunel Jahangirova\textsuperscript{\textasteriskcentered}, 
Fiorella Zampetti\textsuperscript{\textdaggerdbl}, \\
Massimiliano Di Penta\textsuperscript{\textdaggerdbl}, 
Paolo Tonella\textsuperscript{\textdagger}, 
Gabriele Bavota\textsuperscript{\textdagger}}
\IEEEauthorblockA{
\textsuperscript{\textdagger}\textit{Software Institute -- USI Universit\`{a} della Svizzera italiana, Switzerland}\\\textsuperscript{\textdaggerdbl}\textit{University of Sannio, Italy}\\\textsuperscript{\textasteriskcentered}\textit{King's College London, UK}\\
\{rosalia.tufano, nargiz.humbatova, paolo.tonella, gabriele.bavota\}@usi.ch, \\
 d.bifolco@studenti.unisannio.it,
\{fzampetti, dipenta\}@unisannio.it,\\ 
 gunel.jahangirova@kcl.ac.uk}
}

\maketitle

\IEEEdisplaynontitleabstractindextext
\IEEEpeerreviewmaketitle

\input{introduction}
\input{related}
\input{design}
\input{results}
\input{threats}

\input{conclusion}
\input{acknowledgements}

\balance

\bibliographystyle{IEEEtran}
\bibliography{main}

\end{document}

%% file: introduction.tex
\section{Introduction} \label{sec:intro}
Large language models (LLMs) and AI coding assistants have become part of the software development process  \cite{tufano:msr2024}. While these tools may increase productivity \cite{Cui2024Productivity}, they can also produce code that is incorrect \cite{Tambon:emse2025,dou2025arxiv,gao2025arxiv}, insecure \cite{cotroneo:icpc2024,Mousavi:ccs2024,Fu:tosem2025}, inefficient \cite{Abbassi:icsme2025}, and difficult to maintain \cite{Liu:tosem2024}. Recent industry reports also suggest that these risks are not merely hypothetical: according to the Financial Times \cite{amazon-meeting}, internal Amazon material linked a recent trend of high-impact outages in part to ``\emph{Gen-AI assisted changes}'' and to ``\emph{novel GenAI usage for which best practices and safeguards are not yet fully established}''. In this context, understanding the kinds of quality issues that arise in AI-generated code is increasingly important, as it can help in designing new quality assurance practices or adapting existing ones (\eg code review, testing) to better detect issues in AI-generated code. 

Although several studies have characterized quality issues in AI-generated code, most of their evidence comes from evaluating LLMs on code generation benchmarks \cite{Tambon:emse2025,Liu:tosem2024,dou2025arxiv,Abbassi:icsme2025,liu:arxiv2026}. These benchmarks consist of coding tasks defined by (i) a textual description used as the prompt and (ii) a test suite to assess the correctness of the generated code. While this line of work has been essential to improving LLMs' code generation capabilities, it is not intended to identify quality issues in real-world AI-generated code. 

These benchmarks do not capture realistic software development tasks; instead, they focus on relatively simple programming exercises. HumanEval \cite{humaneval}, for instance, has been widely used in prior studies on this topic \cite{dou2025arxiv,gao2025arxiv,Abbassi:icsme2025}. Newer benchmarks such as CoderEval \cite{CoderEval} partly mitigate this issue by deriving tasks from open-source projects and asking LLMs to generate functions from their \texttt{docstring} \cite{Tambon:emse2025,Zhang:issta2025}. However, the prompts used in these evaluations may not be representative of real human-AI interactions in software development. Thus, benchmark-based studies may surface quality issues that differ from those encountered in real development settings. 

A realistic understanding of quality issues in AI-generated code requires evidence from sources that capture how AI tools are used in real-world software development. We address this need by studying developer-attributed quality issues in AI-generated code through three complementary empirical lenses. First, we mine GitHub commits whose messages explicitly indicate that developers are fixing a problem caused by or associated with AI-generated code, identifying \commitsStudyOne of them. We manually inspect all these commits to discard false positives and classify the type of quality issue being corrected at a fine level of granularity (\eg \emph{Functional correctness} $\rightarrow$ \emph{Data parsing and handling} $\rightarrow$ \emph{Hard-coded invalid value/placeholder}). 

Second, we analyze \validConversationsStudyTwo conversations between developers and AI assistants shared on GitHub, totaling \validRoundsStudyTwo message exchanges between the human and the AI. 

\eject

We identify conversations that include source code and textual signals of an LLM error, and then manually inspect them to classify the underlying issue. This conversation-based analysis captures issues detected before the AI-generated code is integrated into the codebase, providing a complementary view of the same phenomenon compared to repository commits. Third, we interview \practitionersStudyThree practitioners and researchers who use AI coding tools, inspecting the transcripts to enrich and validate the taxonomy emerging from repository data. 

Our work has implications for different stakeholders. For researchers, it identifies concrete code generation issues that should be mitigated through improved prompting support, review workflows, and guardrails. For practitioners, it provides a vocabulary for recognizing and discussing the quality risks of AI-generated code, while highlighting specific aspects that deserve particular attention during quality assurance. More broadly, our findings provide a clearer understanding of how to make AI-assisted software development more effective.

All study material is publicly available \cite{replication}.

%% file: related.tex
\section{Related Work} \label{sec:related}




Liu \etal \cite{Liu:tosem2024} run ChatGPT on 2,033 programming tasks from LeetCode. Then, they employ static analysis to analyze the generated code and, through an open card-sort discussion, identify four categories of quality issues: Compilation and Runtime Errors, Wrong Outputs, Code Style and Maintainability, and Performance and Efficiency. Wang \etal \cite{Wang:icse2025} run six LLMs on HumanEval \cite{humaneval} and, via open coding, inspect 557 incorrect solutions to identify the errors they exhibit. They document the root causes behind the errors.  

Tambon \etal \cite{Tambon:emse2025} run three LLMs on CoderEval \cite{CoderEval}, featuring functions mined from GitHub. The input to the LLMs is the \texttt{docstring} of each function. They sample 333 failed code generations and, through open coding, define ten types of quality issues, then validated by surveying 34 participants.

Other similar works adopted a (semi-)automated approach to classify the quality issues in AI-generated code. Liu \etal \cite{Liu:tse2024} run ChatGPT on 728 LeetCode problems and inspect 157 wrong answers with the goal of classifying the changes required to address the bug (\eg Single-hunk $\rightarrow$ Add Statements). They show that most bugs require large fixes. Dou \etal \cite{dou2025arxiv} ran nine LLMs on three benchmarks 
and automatically created a taxonomy of bugs by relying on the compiler feedback. Then, 22 human experts examined the incorrect code and annotated the bug type, resulting in a taxonomy of ten bug types. 
Abbassi \etal \cite{Abbassi:icsme2025} ran three LLMs on HumanEval+ \cite{evalplus}, and used GPT-4o-mini as a judge for labeling inefficiencies in the 492 generated code snippets. Open coding has then been used to categorize the inefficiencies in five categories: General Logic, Performance, Readability, Maintainability, and Errors. 

While prior studies mainly characterize AI-generated code issues through controlled benchmark evaluations, we examine quality issues documented in real software development practice. 

\eject

The commit- and conversation-based analyses allow to study problems that arise in authentic use cases, where developers interact iteratively with AI tools, integrate generated code into evolving projects, and discover issues through review, testing, and debugging. The need for more realistic studies was also highlighted in the literature review by Gao \etal \cite{gao2025arxiv} about quality issues in LLM-generated code. They conclude that ``\emph{most existing studies focus on relatively simple code-generation tasks, with limited attention to more complex, context-dependent development scenarios}''. As detailed in \secref{sec:comp-existing}, our taxonomy includes 50 AI-generated quality issue types undocumented in the existing taxonomies.

Other studies focus on specific types of quality issues, such as simple (one-liner) bugs \cite{Jesse:msr2023}, security-related issues \cite{cotroneo:icpc2024,Mousavi:ccs2024,Fadi:ceur,Fu:tosem2025}, and hallucination-related issues \cite{liu:arxiv2026,Zhang:issta2025}. 

Moratis \etal \cite{Moratis:msr2024} examine 453 developer-AI conversations running the PMD static analysis tool on them,  detecting violations relevant to standard practices, code readability, and bugs. Siddiq \etal \cite{Siddiq:msr2024}, instead, use static analysis to detect code and security smells in conversations. The manual inspection of conversations allowed us to identify quality issues beyond those detectable by static analyzers (\eg wrong test oracles).

%% file: design.tex
\section{Study Design} \label{sec:design}

Our study answers the following research question: \textbf{\emph{What types of quality issues do practitioners attribute to AI-generated code in real-world software development?}} 



\subsection{Study Context \& Data Collection}
Our study is based on three complementary sources of evidence: repository artifacts, shared developer-AI conversations, and practitioners' accounts collected through interviews. Together, these sources allow us to investigate both quality issues that were integrated into real projects and later fixed, and issues identified earlier during interactions with AI tools. 

\subsubsection{Commits Fixing AI-generated Code}
We rely on the GHArchive dataset \cite{GHArchive}, which records public GitHub activity in hourly \texttt{json} archives. 
We extract all commits performed between Jan 1, 2022, and Oct 31, 2025. We chose the start date to fall approximately two months after the release of the GitHub Copilot technical preview (Oct 29, 2021). Copilot can be regarded as the first LLM-based code generation tool deployed at scale. Its release was followed by several other AI coding tools, including general-purpose LLMs such as ChatGPT (Nov 2022). Regarding the end date, from November 2025, GHArchive no longer provides event details.

We then defined a set of lexical patterns to match commit messages documenting the resolution of quality issues introduced by AI-generated code. The patterns combine four categories of lexical signals: corrective actions (\eg fixing, patching), problem indicators (\eg bug, issue, error), causal-attribution phrases (\eg introduced by, caused by, due to), and mentions of AI coding tools or LLMs (\eg Copilot, ChatGPT, Claude). Examples of matched commits include ``\emph{fixed bug introduced by Copilot}'', ``\emph{corrects silly mistake by GPT-4}''. 

Requiring all four components reduces ambiguity and increases the likelihood that retrieved commits truly document fixes to AI-introduced quality issues. We constructed the pattern iteratively. We started from a small set of seed terms (\eg all commits featuring ``bug'' and ``by Copilot''), inspected the retrieved commit messages, and then expanded the pattern with lexical variants and alternative spellings observed in practice. At the same time, we excluded terms that systematically introduced noise (\eg commits including the term ``review'' often referred to the fixing of issues recommended in a review generated by an AI). We also introduced bounded proximity constraints between the four components to favor commit messages in which the fixing action, problem description, causal attribution, and AI-tool reference occur in close textual proximity, reducing spurious matches arising from unrelated co-occurrences. The proximity thresholds were refined iteratively by inspecting preliminary matches and balancing retrieval breadth against noise. For example, we allow at most 20 characters between the causal-attribution phrase (\eg \emph{caused by}) and the AI coding tool (\eg \emph{Copilot}). This means that we match commit messages such as ``\emph{fixing bug caused by code generated by Copilot}'', while avoiding those in which ``\emph{caused by}'' and ``\emph{Copilot}'' appear quite far. 

The list of AI tools targeted by our patterns was derived from recent surveys of LLMs and AI agents used for software-related tasks \cite{robbes2026promises}. Overall, our retrieval covers \assistantsStudyOne AI-based tools. When necessary, we defined multiple lexical variants for the same tool to account for alternative names and usages in commit messages. For example, our patterns include GitHub Copilot, Copilot, Copilot Studio, Copilot M365, and Codex (\ie the model underlying Copilot). The Python scripts implementing the complete regexes are publicly available \cite{replication}. Importantly, these regexes were used only to retrieve candidate commits for manual inspection, and not to automatically classify commits as true instances. Irrelevant matches were filtered out during the subsequent manual analysis (\secref{sub:manualAnalysis}). 

This process led to \commitsStudyOne commits for manual inspection.

\subsubsection{Shared Developer–AI Conversations on GitHub}
Xiao \etal \cite{Xiao:MSR2024} recently introduced DevGPT, a dataset featuring developer-ChatGPT conversations linked to software artifacts. Our initial plan was to use this dataset to identify conversations in which the LLM apologized for mistakes in generated code, and then manually inspect those conversations to classify the quality issue referenced in the apology. However, DevGPT only includes conversations up to 2023, thus excluding developers' interactions with newer LLMs. For this reason, we replicated the methodology from Xiao \etal to build an updated version of their dataset, which we also release as an additional contribution of our work \cite{replication}. 

We used the GitHub Search APIs \cite{GitHubSearch} to search for files, commits, issues, and PRs that contain at least one of the following links: \texttt{https://chatgpt.com/share/} or \texttt{https://chat.openai.com/share/}, which are used to share ChatGPT conversations. Unlike Xiao \etal, we did not collect links from Hacker News articles or GitHub discussions, as we considered them less relevant to our study. 

We then parsed the collected software artifacts to extract all relevant links they featured, and downloaded the shared conversations. We also tried to download conversations shared by Claude and Gemini. However, despite multiple attempts, every crawler we built was blocked after collecting only a few conversations.

After removing duplicates and conversations not accessible anymore, we ended up with \linksStudyTwo valid conversations (\filesStudyTwo from files, \commitsStudyTwo commits, \issuesStudyTwo issues, and \prsStudyTwo PRs). 
One clarification is due regarding the completeness of the retrieval process. For commits, issues, and PRs, our search was exhaustive, meaning we collected all results returned by the GitHub APIs. This was not possible for files. Indeed, the API returns at most 1,000 file results per query. Thus, the set of shared conversations we retrieved from the files is limited to those in the first 1,000 matching files. We considered issuing the query multiple times to extend coverage; however, the file results are neither returned nor filterable by a stable sorting criterion. This prevents a reliable partitioning of the search space and, in turn, makes exhaustive collection infeasible. This is not a significant limitation for our study, since we target manual inspection of a sample of such conversations. 

To increase the likelihood of inspecting relevant conversations, we automatically filtered out conversations with characteristics unlikely to provide useful evidence. First, we excluded conversations consisting of only one message exchange between the developer and the AI (\ie one user message and one LLM response), since these cannot contain textual signals of LLM mistakes: they include only the initial request and the corresponding answer, with no opportunity for the user to report an error. We also excluded excessively long conversations, as we observed that they often correspond to ``tangled conversations'' spanning multiple unrelated tasks, making it difficult during manual analysis to reconstruct the context of each message. To set an upper threshold, we computed the number of message exchanges for each conversation and applied the standard outlier criterion Q3 + 2.5$\times$IQR, where Q3 denotes the third quartile and IQR the interquartile range. This yielded a threshold of 19 exchanges, so we removed all conversations longer than 19 rounds (38 messages), leaving 3,909 conversations. We then removed conversations without any code snippets, as these were unlikely to concern code generation. Finally, to focus on conversations in which the LLM likely made a mistake, we searched for LLM responses containing apology-related textual patterns (\eg ``\emph{You're right!}'', ``\emph{I apologize}''). We derived these patterns by first asking GPT 5.2 for sentences it commonly uses to apologize for mistakes. Then, we generalized the resulting list to account for both American and British English and expanded it after inspecting a sample of 30 conversations. For example, ``\emph{You're right!}'' was not suggested by ChatGPT but added by us. The script used for this filtering step is publicly available \cite{replication}. After all filters, we obtained \validConversationsStudyTwo candidate conversations, manually inspected as described in \secref{sub:manualAnalysis}.

\eject

\subsubsection{Developer Interviews}
We targeted two groups of developers who regularly use AI code generators: researchers and practitioners. 
We recruited participants through personal contacts, following a convenience sampling strategy \cite{Etikan2016ConvenienceSampling}. This yielded 42 candidates that were contacted by email. The invitation described study goals, the expected interview duration (30 minutes), and the voluntary nature of participation. Also, it informed invitees that they would be asked to consent to (i) the interview being transcribed for analysis, (ii) the transcript being shared in anonymized form, and (iii) the findings being published in a research article.
We received 24 positive responses, of which 20 resulted in scheduled interviews: 3 of the participants are active researchers, and 17 are practitioners with work experience between 4 and 25 years (mean = 9.6, median = 8.5) . The reported experience with AI-assisted coding tools ranged from 0.5 to 5 years (mean = 2.75, median = 2.5). 


All the interviews were conducted online using MS Teams or Google Meet. They lasted between 26 and 73 minutes (mean = 45 minutes). The interviews were automatically transcribed, and two of the authors manually reviewed and corrected the transcripts when needed.

The interview goal was not to validate the taxonomy categories extracted during the mining study but to avoid bias by letting respondents naturally report the problems they encountered. Therefore, we adopted a \textit{semi-structured interview} design \cite{Seaman:1999}, combining open-ended questions, to elicit unexpected insights, with more focused ones, to keep the discussion within scope. Because this format often requires improvising follow-up questions, each interview was conducted by two authors with complementary roles: one led the interview, while the other intervened with additional questions when appropriate. After collecting background information on participants' experience with AI-assisted coding, we proceeded with the questions from our interview guide \cite{replication}. Our first question examines participants' overall experience and the challenges they face when producing and working with AI-generated code: ``\emph{Can you describe your experience using AI coding tools in your development work, particularly in terms of the quality of the code they generate and any challenges or issues you have encountered?}''. The next questions prompt the provision of specific examples and reasons for any problems encountered. We then proceed to a block of more specific questions inspired by the top categories of our taxonomy. We did not provide interviewees with any particular sub-category or leaf examples. A representative question from this part of the interview is ``\emph{Have you experienced any issues related to the performance of AI-generated code?}''. After becoming familiar with the issues our participants experienced, we asked whether they thought these faults were similar to those found in human-written code. We then asked whether participants ever found themselves in a situation where they had to give up on the generated code, and why. The final part of the interview focuses on practices of detecting, handling, and fixing faults in AI-generated code. 

\subsection{Coding Procedure}
\label{sub:manualAnalysis}
We describe the manual analysis procedures applied to \emph{fixing commits}, \emph{developer-AI} conversations, and \emph{interview transcripts} to identify the types of quality issues in AI-generated code.

\subsubsection{Commits Fixing AI-generated Code}
We manually inspected all \commitsStudyOne candidate commits fixing quality issues caused by AI-generated code using open coding~\cite{Seaman:1999}. Each commit was assigned to two assessors, with a total of five authors involved. Each assessor independently inspected the commits assigned to them. The first goal of the manual analysis was to determine whether each candidate commit was a \emph{true} or a \emph{false positive}. We classified a commit as a \emph{true positive} when there was strong evidence that the developer was fixing a quality issue introduced by AI-generated code. Examples of true positives include commits accompanied by messages such as ``\emph{fix syntax errors caused by me trusting Copilot}'' and ``\emph{fixing issues due to Claude}''. In contrast, some cases were straightforward false positives, such as ``\emph{fix issues reported by Copilot}''. Other cases were more ambiguous. For example, the message ``\emph{fix bugs by Cursor}'' could refer either to bugs introduced or detected by the AI. In such situations, we examined the repository context to identify evidence on whether the tool had been used for code generation or code review. When no clear evidence supported either interpretation, we conservatively classified the instance as a \emph{false positive}. We also excluded duplicate commits due to forked repositories.

For each instance classified as \emph{true positive}, the assessor assigned one or more labels describing the quality issue(s) fixed by the developer in the commit. We did not begin with a predefined set of labels to avoid biasing the analysis; instead,  labels emerged from the data. The guideline for the assessors was to assign labels as precise as possible favoring, for example, a label in the form ``\emph{Build / compilation / static errors $\rightarrow$ Name/interface resolution $\rightarrow$ Non-existing function/attribute}'' rather than generic labels as done in previous work: ``\emph{Compilation and Runtime Errors}'' \cite{Liu:tosem2024}, ``\emph{Syntax Error}'' \cite{Tambon:emse2025}, ``\emph{Hallucination}'' \cite{dou2025arxiv}, and ``\emph{Error}'' \cite{Abbassi:icsme2025}. The informativeness of our labels is  considered key for practical applications of our findings (\eg customizing code review practices). We also introduced a special ``\emph{Unclear}'' label for cases in which the fixed quality issue could not be determined with confidence.



The open coding of commits was conducted by following a reflexive thematic analysis approach without predetermined categories \cite{braun2006using}.
It was supported by a shared Google Sheet in which the authors progressively recorded the labels they introduced. Although, in principle, this is against the notion of open coding, it was necessary to prevent the label space from growing excessively and to promote consistent naming across assessors, while still allowing categories to emerge from the data. After an initial round of labeling on 100 commits, the involved authors held an alignment meeting to discuss all disagreements, review and consolidate the labels defined up to that point, and refine the coding guidelines. We then proceeded with two additional batches of 100, introducing new labels as needed and consolidating them at the end of each batch. 

After these first three rounds, we proceeded to inspect all remaining 614 commits, keeping the possibility of adding new labels open. Conflicts were resolved at the end of each round, considering that, being this a multi-label categorization, in some cases, labels from both annotators could be considered valid. Whenever the two original evaluators disagreed --- either on the true/false positive classification or on the labels assigned to describe the fixed quality issue --- the commit was reviewed by two additional authors who had not been involved in the initial assessment of that commit.

We determined the reliability of false positive classification using Cohen's $k$ \cite{kappa} inter-rater agreement, which resulted in $k$=0.76 (substantial agreement). For what concerns the categorization, since we  performed a  reflexive thematic analysis  without predetermined categories, with two different annotators possibly assigning two valid codes to the same commit (\eg one related to functional and one to nonfunctional aspects), inter-rater agreement is not recommended \cite{braun2006using}.


Out of the \commitsStudyOne inspected commits: 131 were not accessible anymore, 68 were duplicates, 485 \emph{false positives}, and \relevantCommitsStudyOne \emph{true positives}. The \emph{true positives} were used to derive a first version of our (hierarchical) taxonomy that featured 51 types of quality issues. The taxonomy was derived through three iterations involving five authors, who discussed how to cluster the leaves into intermediate and root nodes and, when needed, how to group very closely related leaves.


\subsubsection{Shared Developer–AI Conversations on GitHub}
We manually inspected all \validConversationsStudyTwo candidate developer-AI conversations following the same procedure adopted for commits. Each conversation was assigned to two assessors, totaling eight authors involved in this analysis. 
The collected conversations contained a total of \validRoundsStudyTwo message exchanges. We conducted an initial alignment round across 55 conversations (456 message exchanges) to refine the coding guidelines for this specific type of artifact. 
We then proceeded with the remaining 244 conversations (1,407 message exchanges).

Each assessor independently inspected the conversations assigned to them. As in the commit analysis, the first goal was to determine whether each message exchange was a \emph{true positive} or a \emph{false positive}. A \emph{true positive} provides strong evidence that either the LLM is apologizing for a mistake made during code generation or the user is explicitly pointing out such a mistake, and the error is visible in the code produced in the previous message exchange. All other cases were classified as \emph{false positives}. For each \emph{true positive}, the assessor assigned one or more labels describing the quality issue(s) under discussion, following the same coding procedure used for commits. In this case, however, we initialized the coding process with the 51 quality issue types identified in the commit-based analysis, extending the label set whenever needed. 
To support the analysis, we developed a web application in Streamlit \cite{Streamlit}, to display the conversations assigned to each assessor. The tool is publicly available \cite{replication}. 

As in the commit-based analysis, disagreements were resolved through additional assessment. Whenever the two original evaluators disagreed --- either on the true/false positive classification or on the assigned labels --- the round was reviewed by two additional authors not involved in its initial assessment. The Cohen's $k$ inter-rater agreement for false-positive classification was $k$=0.70 (substantial agreement).

At the end, we identified 453 message exchanges as \emph{true positives} (1410 \emph{false positives}). The analysis of the \emph{true positives} extended the taxonomy with 25 new leaf nodes (\ie quality issue types not observed in the commit-based analysis).

\subsubsection{Interview Transcripts}
We  open-coded the \practitionersStudyThree interview transcripts. Each transcript was assigned to both interviewers, who independently labeled passages in which an interviewee mentioned quality issues encountered while coding with AI assistance. We initiated coding with the 76 issue types identified in the commit and conversation analyses, and expanded the set as needed. Overall, 269 text passages were tagged by at least one evaluator. A text passage is a contiguous segment of the transcript in which the interviewee discusses a specific type of quality issue. Passages averaged 92 words (median: 65). In 93 cases, a passage was tagged by only one evaluator; in 73 cases, the two evaluators assigned different labels to the same passage. Disagreements were resolved using the same procedure adopted for commits and conversations, yielding a final set of 185 valid passages labeled with a quality issue. The number of labeled passages per interview ranged from 4 to 16, with a mean of 9.25. 

This process further enriched our taxonomy with 11 leaf nodes and one sub-category. The final taxonomy is shown in \figref{fig:taxonomy}. \tabref{tab:new_categories_per_round} reports the number of new labels introduced at different taxonomy levels (\ie root, inner, and leaf nodes) across the seven inspection rounds. For each round, the table also reports: (i) the type of artifact inspected, \ie commits, conversation rounds, or  interview answers with at least one sentence containing more than 3 words (to exclude uninformative answers); (ii) the number of inspected instances of that artifact type (\eg 100 commits in the first round); and (iii) the average number of instances that had to be inspected to identify a new leaf node.
The latter metric reveals an interesting pattern: later rounds required inspecting more instances to identify a new label. For example, in the commit-based analysis, we moved from one new leaf node every 5.9 inspected commits in the first round to one every 36.1 commits in the fourth round. The last round of conversation analysis required reading an average of 87.9 message exchanges (rounds) between the developer and the AI to find a new label. When it comes to interviews, the results show that we had to analyze on average 220.8 interview answers to discover one new leaf.  

\begin{table}[h]
	\centering
	\scriptsize
	\caption{New quality issue types per inspection round.}
	\label{tab:new_categories_per_round}
	\begin{tabular}{clrrrrr}
		\toprule
		\multirow{2}{*}{\textbf{Round}} & \multirow{2}{*}{\textbf{Artifacts}} & \# \textbf{Instances} & \textbf{New} & \textbf{New} & \textbf{New} & \textbf{\# instances}\\
		& & & \textbf{root} & \textbf{inner} & \textbf{leaf} & \textbf{per new leaf}\\
		\midrule
		1 & commits       & 100 & 4 & 11 & 17 & 5.9  \\
		2 & commits       & 100 & 3 & 8  & 11 & 9.1  \\
		3 & commits       & 100 & 0 & 1  & 6  & 16.7 \\
		4 & commits       & 614 & 1 & 2  & 17 & 36.1 \\
		5 & conv. rounds & 456  & 0 & 1  & 9  & 50.6  \\ 
		6 & conv. rounds & 1,407 & 0 & 1  & 16 & 87.9 \\ 
		7 & int. answers    & 2,429 & 0 & 1 & 11 & 220.8 \\
		\bottomrule
	\end{tabular}
\end{table}

%% file: results.tex
\section{Results Discussion} \label{sec:results}

\begin{sidewaysfigure*}[p]
	\centering
	\vspace*{\fill}
	\hspace{-0.5cm}
	
	\includegraphics[
	width=\textheight,
	keepaspectratio
	]{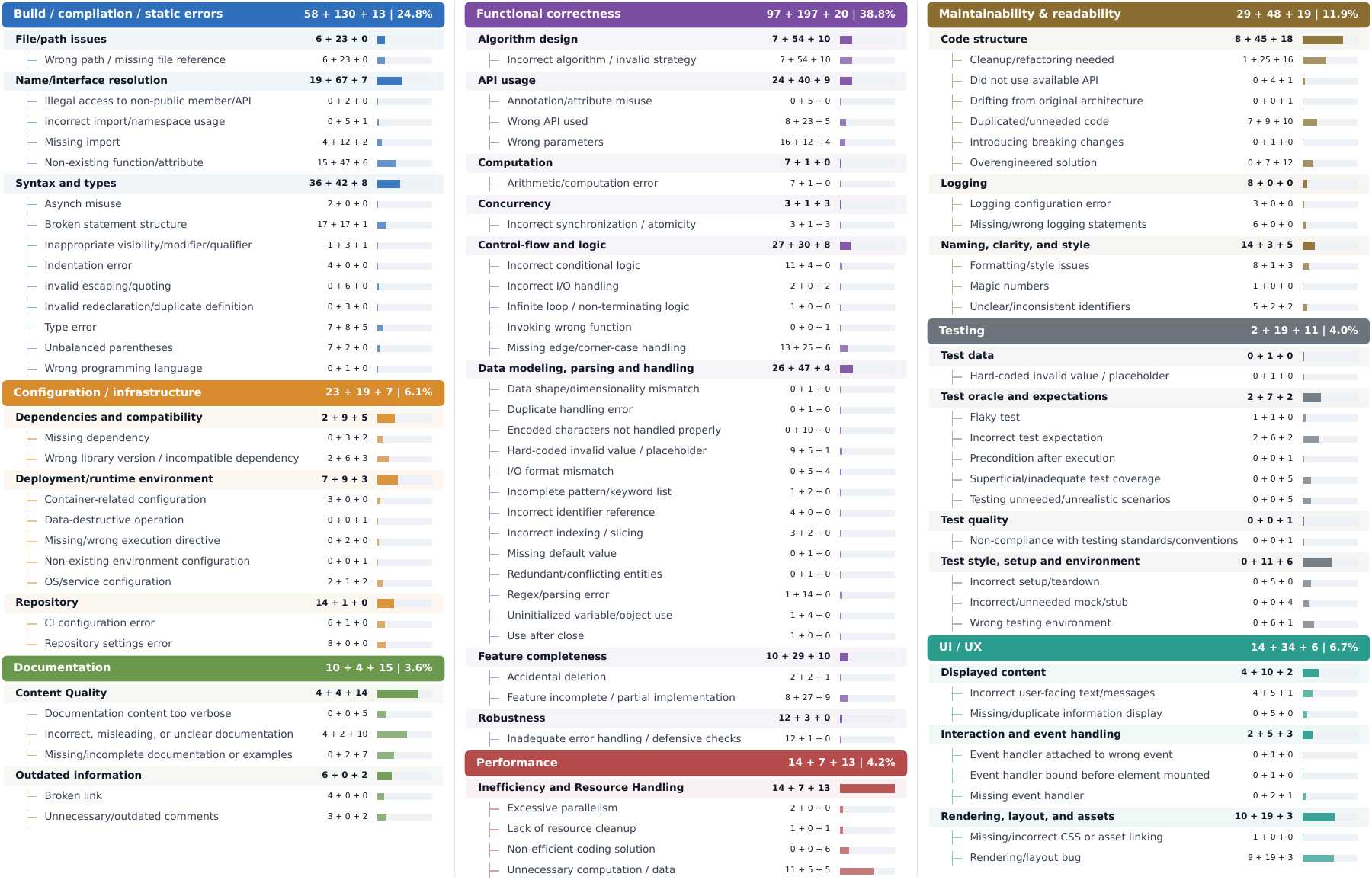}
	
	\caption{Taxonomy of quality issues in AI-generated code.}
	\label{fig:taxonomy}
	
	\vspace*{\fill}
\end{sidewaysfigure*}

Our taxonomy (\figref{fig:taxonomy}) comprises eight root categories, each further organized into hierarchical subcategories. For each category, the three numbers reported after its name (separated by plus signs) denote the number of supporting instances identified in commits, developer-AI conversations (message exchanges), and interviews, respectively. The taxonomy does not show the \emph{Unclear} category, which contains 33 inspected artifacts from commits and conversations. 
In the following, we discuss each root category, reporting qualitative examples of issues we found and highlighting lessons learned with the $\blacktriangleright$ icon. Due to limited space, we do not discuss all issue types. The interested reader can find all instances with the corresponding labels in our replication package \cite{replication}.

\subsubsection*{\textbf{Functional Correctness}} Unsurprisingly, this is the largest category in our taxonomy, accounting for 38.8\% of true positive instances and 27 of the 87 leaf nodes (31\%). A prominent subcategory is \emph{Data modeling, parsing and handling}, with 77 instances and 13 leaf nodes. One frequent issue is the introduction of \emph{hard-coded invalid values/placeholders}, especially for \texttt{String} values required by the implementation that cannot be inferred by the model. For example, in one inspected commit \cite{https://github.com/Stensel8/Scripts/commit/1e270a21805997353c9ca038a89770e9a0f8cd46}, the developer wrote: ``\emph{Correcting a mistake caused by Copilot. Changed the hash to the actual value, it should be}''. Copilot updated the OpenSSL library from 3.5.1 to 3.5.2 and also changed the \texttt{OPENSSL\_SHA256} constant to an invalid hash. Since this constant is used to verify the downloaded archive, the script would have downloaded the correct OpenSSL version but rejected it during checksum verification. Thus, the AI correctly updated the dependency version, but hallucinated an invalid checksum. $\blacktriangleright$ This suggests that generated values that cannot be derived from the local context (\eg hashes, tokens, URLs) should be (i) specified in the prompts when possible; and (ii) explicitly verified by developers inspecting the AI-generated code.

Other frequent issues in this subcategory include \emph{regex/parsing errors} and \emph{improper handling of encoded characters}. In one conversation \cite{https://chat.openai.com/share/b78b8a63-fb6e-41c2-a874-3c682b61578f} the developer used ChatGPT to extract receipt information from images. Instead of designing a robust parser, the model repeatedly patched a regex to handle newly observed receipt-line formats. These patches relied on assumptions about token positions and mandatory fields, and later introduced unsafe optional groups causing runtime errors. $\blacktriangleright$ This illustrates how AI-generated parsers may overfit to the examples seen during interaction, producing plausible patches while failing to capture the variability of real-world inputs.

LLMs also struggle to devise correct implementation strategies, as shown by the 71 instances in \emph{Algorithm Design} and the 65 in \emph{Control-flow and logic}, with one fault type of the latter (\emph{invoking wrong function}) coming exclusively from the interviews. \emph{Algorithm Design} captures cases in which the model adopted an unsuitable or fundamentally incorrect approach, while \emph{Control-flow and logic} covers plausible solutions affected by localized logic errors. 

\eject

One algorithmic error comes from commit \texttt{541c696} of the \texttt{rankify-api} repository \cite{https://github.com/Tonypsilon/rankify-api/commit/541c6962606f21a13ed2d74686e0d64619501901} in which the developer documented: ``\emph{Fix ordering issue introduced by Copilot. Was sorting poll options alphabetically for no reason.}'' Similarly, in one inspected conversation \cite{https://chatgpt.com/share/68121cb1-e234-8005-a86b-5f924a0f7784}, the AI failed on a text post-processing task requiring the restoration of stress markers (\ie symbols such as a backtick indicating which syllable or vowel is stressed). The generated solution treated the marker as punctuation and separated it from the word, splitting stressed words into multiple tokens and preventing subsequent logic from restoring the marker correctly. Instead of addressing this root cause, the AI repeatedly patched spacing, punctuation, and word matching, without fixing the underlying algorithmic error.  $\blacktriangleright$ This illustrates a recurring failure mode: the model may keep repairing visible symptoms while missing the underlying abstraction or algorithm. Algorithmic errors appeared far more often in developer-AI conversations (54 instances) than in commits (7 instances), likely because developers rarely commit implementations whose overall logic is visibly wrong. At the same time, this type of issue was mentioned by 45\% of interview participants. $\blacktriangleright$ For researchers, this suggests that repository histories alone may underestimate high-level reasoning failures in AI-generated code.

A frequent localized logic error is the failure  \emph{Missing edge/corner-case handling} (44 instances). In commits, this often appears as a small fix, such as adding a \texttt{null} check. In conversations, it typically follows a recurring pattern: the user requests an implementation, tests the generated code, reports an error, and the AI acknowledges that a specific case was not handled. For example, in one conversation \cite{https://chat.openai.com/share/6c356c82-3607-4c92-8271-99144239b1ae}, the developer reported `\texttt{firstBatch.items is not iterable}'', and the AI replied: `\emph{I apologize for the oversight [$\dots$] let's add a check to ensure that the items property exists and is an array before proceeding}''. $\blacktriangleright$ Developers should explicitly check whether AI-generated code assumes ideal input structures and whether it handles missing, null, or malformed data.

\emph{API usage} also appears to be problematic for LLM-generated code. The most common cases involve either selecting the wrong API to implement a feature (36 instances) or invoking the correct API with wrong parameters (32 instances). An example of the former appears in commit \texttt{ff32d75} of the \texttt{TypeAgent} repository \cite{https://github.com/microsoft/TypeAgent/commit/ff32d75e4cbd80726549a2f0bfaaa9b4ea677886}, whose message states: ``\emph{Fix bugs introduced by Copilot (and my poor review skills). I thought it had the type wrong when calling selectRelevantChunks(), but actually it got the name wrong, and it should have called selectChunks().}'' These cases differ from invocations of non-existing functions, which we classify under \emph{Build/compilation/static errors}; here, the API exists, but it is semantically inappropriate for the intended task.

Errors involving \emph{wrong parameters} often manifest as user-visible functional bugs, and may stem from the model mixing conventions across libraries that offer similar functionality. For example, in one inspected conversation \cite{https://chatgpt.com/share/6713d47d-4960-8003-9869-293729ea5b8c}, the user reported a bug in the generated code, and the LLM replied: ``\emph{I see the issue now. The keyboard.is\_pressed(20) is mistakenly detecting the key code for the T key instead of Caps Lock, because in the keyboard library, Caps Lock doesn't directly map to its typical key code}''. $\blacktriangleright$ Developers should review AI-generated API calls not only for type compatibility or existence, but also for semantic fit. For researchers assessing code generators, these cases highlight the need for evaluations that go beyond compilation and check API-level semantic correctness.

Finally, \emph{Feature completeness} includes 50 instances in which the AI-generated code failed to implement all required functionality. These cases appear mainly in developer-AI conversations and were experienced by 50\% of interviewees. They are often caused by the model losing track of requirements across multi-turn interactions. A concrete example comes from a conversation about building an F\# program \cite{https://chat.openai.com/share/821d8f25-f37c-4cfc-8ca5-0ff7200e3f04}. After receiving an initial implementation, the developer asked the model to add ``\emph{a Yield method that sets all the values to reasonable defaults}'', which the model successfully implemented. The developer then asked to extend the implementation to additional data types. However, while extending the code, the model removed the previously requested and correctly implemented \texttt{Yield} methods, prompting the developer to reply: ``\emph{You left out all the Yield methods}''. $\blacktriangleright$ In multi-turn development, developers should verify not only whether newly requested functionality is implemented, but also whether previously satisfied requirements are preserved. Researchers need to evaluate AI code generation capabilities for requirement retention across iterative interactions, rather than only for single-turn task completion.

\subsubsection*{\textbf{Build/compilation/static errors}} This is the second largest root category in our taxonomy, accounting for 25.2\% of true positive instances and 14 leaf nodes (16.1\%). We discuss three main subcategories. The first is \emph{File/path issues} (31 instances), which captures cases where the generated code contains incorrect paths or file references. These errors may not only prevent files from being found but also cause generated command invocations to fail. For example, in commit \texttt{9e4b69e} of the \texttt{tether-router-monitor} project \cite{https://github.com/leonzdev/tether-router-monitor/commit/9e4b69e695fe2c40e8794b1a9cab4a18b324163a}, the developer fixed ChatGPT-generated code that invoked shell commands using an incorrect path, such as \texttt{executeShellCommand("./ifdev")} instead of \texttt{executeShellCommand("ifdev")}. $\blacktriangleright$ AI-generated paths and command invocations should be checked against the actual project layout and runtime environment. 

The second subcategory concerns \emph{name/interface resolution} errors (94 instances), most of which involve \emph{non-existing functions or attributes} (69 instances). A concrete example comes from a conversation we inspected \cite{https://chatgpt.com/share/d85c4695-4b71-4e13-a4b2-51f14031198f}, in which the model invoked \texttt{getElement}, a function not available on the target type \texttt{Exprs::Tuple}. After the developer reported the error, the assistant acknowledged the mistake but proposed several additional problematic fixes, leading to further name-resolution errors. After a few interactions, the model even returned to a solution exhibiting the original problem, namely ``\emph{getAnElement() cannot be resolved for type Exprs::Tuple}''. In this case, the model repeatedly fabricated API methods while maintaining a confident explanatory style. 

As a result, it produced plausible-looking code that failed at type checking and shifted the burden of API validation to the developer. $\blacktriangleright$ Developers should treat AI-generated calls to unfamiliar APIs as hypotheses to be verified. Researchers, again, should evaluate LLMs in multi-turn conversations.

Finally, \emph{syntax and type} errors are also frequent in AI-generated code (87 instances), making this the second largest inner category in the whole taxonomy after \emph{name/interface resolution}. Most of these errors involve \emph{broken statement structures}, where the generated code violates the grammar of the target language. For example, in a commit fixing ``\emph{syntax issues from using gemini}'' \cite{https://github.com/Ztgunderson/ECE455/commit/211808c1b4c385f42de4b17f1de791e8f060a150}, the model produced an invalid CUDA kernel declaration, using \texttt{\_\_global} instead of \texttt{\_\_global\_\_}. $\blacktriangleright$ Such errors suggest that models may approximate unfamiliar language extensions or framework-specific annotations exploiting surface similarity to a known language (in the discussed case, Python). This makes specialized toolchains especially prone to invalid code.


\subsubsection*{\textbf{Maintainability \& readability}} LLM-generated code can be syntactically and functionally correct while exhibiting maintainability and readability issues, which account for 11.8\% of instances and 11 leaf nodes (12.6\%) in our taxonomy. This category was mentioned in 95\% of interviews. Most maintainability issues (71 instances) concern \emph{wrong code structure}, often requiring \emph{cleanup/refactoring}.  Some of the involved conversations focus entirely on refactoring \cite{https://chatgpt.com/share/686a0646-7e1c-8002-a487-154858aa3157,https://chatgpt.com/share/686e2bdc-f6c4-8008-9e00-06378c2bdbc1}: the user asks the LLM to improve the structure of existing code, but the model fixes some smells while introducing others, such as removing a ``long parameter list'' smell while introducing a ``long method'' smell \cite{https://chatgpt.com/share/686e2bdc-f6c4-8008-9e00-06378c2bdbc1}. $\blacktriangleright$ AI-assisted refactoring should not be validated only by checking whether the targeted smell disappeared; developers should also verify whether the transformation introduced new structural problems elsewhere.

Developers also frequently flagged \emph{duplicated code} (26 instances), with 50\% of interview participants mentioning this as a persistent issue in AI-generated code. Some cases are trivial, such as Copilot duplicating the same statement in two consecutive lines of a Python function \cite{https://github.com/s3bc40/moccasin/commit/461ad6998a9d4ba0b94a27ac3684e356edeb04c5}, or defining the same class twice in the same file \cite{https://github.com/podaac/bignbit/commit/f68fceb763ae7c24fa6f7439211fe739c580577e}. In other cases, duplication reflects a lack of project-level context: the LLM reimplemented functionality that already existed elsewhere, as in a conversation where the developer noted that ``\emph{PromptDescriptorViewer got reimplemented in TasksLists, but it should be reused instead}'' \cite{https://chat.openai.com/share/e9f4664e-50a5-40c1-8604-befe89a2dd36}. $\blacktriangleright$ When using LLMs, developers should be more careful than ever to monitor the presence of duplicated code, as models may fail to recover relevant project context or to perform abstractions properly. 


\subsubsection*{\textbf{Other root categories}} Due to space constraints, we discuss the remaining five root categories together: \emph{UI/UX}, \emph{Configuration/infrastructure}, \emph{Testing}, \emph{Performance}, and \emph{Documentation}. Rather than covering them exhaustively, we focus on quality issues that, based on our experience, appear particularly informative for AI-assisted development.

Out of 51 instances of the \emph{Configuration/infrastructure} category, 17 artifacts discuss quality issues related to \emph{dependencies and compatibility} that are often difficult for the AI to fix. 

One example comes from a conversation about Go \cite{https://chatgpt.com/share/c4ca8622-9b01-4189-9e29-7414ddf79238}. The AI generated a coding solution relying on an external library, which in Go must be managed through a module configuration file. The AI repeatedly forgot or mishandled this dependency setup. $\blacktriangleright$ For dependency-sensitive tasks, developers should consider more robust interaction patterns that require the AI to generate a complete, runnable unit, including the dependency manifest, working directory assumptions, and the exact execution command, so that dependency errors become explicit rather than hidden in the surrounding setup.

\emph{Performance issues} appeared in both commits and conversations, mostly as \emph{unnecessary computation or data} introduced by AI-generated code. This category is featured in 65\% of the interviews.
In one inspected conversation \cite{https://chat.openai.com/share/608ed300-6714-4c74-be6e-022e80b05207}, AI was used to configure Tailwind CSS, a CSS generator, but the wrong configuration made the tool unnecessarily scan the whole project rather than the HTML source only.

The \emph{testing} root category (32 instances) covers issues with \emph{test oracles and expectations}, \emph{test quality}, \emph{test style, setup, and environment}, and constitutes the biggest contribution from the interviews (1 sub-category and 5 leaf nodes). This reveals that testing activities are underrepresented in the other taxonomy sources, possibly because they are less ``admitted'' in software repositories, yet very relevant for developers.  
Some of the interview participants mention that AI tools significantly lighten the burden of test case implementation (interviews \# 1, 3, 7, 8, 12, 16, 20 \cite{replication}). At the same time, the generated tests are often naive (\# 3, \emph{Testing unneeded/unrealistic scenarios}) and miss expected testing targets (\# 17, 8, 12, 16, \emph{Superficial/inadequate test coverage}). One participant specifically mentions that AI-generated tests degrade sharply as code complexity increases (\# 1). In the majority of cases coming from commits and conversations, the LLM produced tests with \emph{incorrect expectations}. One reason was the loss of context in long interactions. For example, in a 20-round conversation \cite{https://chat.openai.com/share/b0dabec7-a903-4d50-b300-89801ec8e31d}, the LLM was asked to generate tests and replied: \emph{To write unit tests for overlyComplexFunction.js, I'll need to make some assumptions [$\dots$] as I don't have the implementation details.}'' This led to wrong and incomplete tests. The developer then pointed out the inconsistency: \emph{You wrote overlyComplexFunction.js earlier}'', showing that the model had lost track of information available earlier in the conversation. 

Finally, for quality issues related to \emph{UI/UX} and \emph{Documentation}, we only highlight their presence across all types of inspected artifacts, with interviews confirming their relevance for developers. We point the interested reader to our replication package \cite{replication} for concrete examples.



\subsection{Interview Insights} \label{sec:interview-results}

The interviews added 11 new fault types to the taxonomy and confirmed the relevance of all 8 root categories, 21/26 inner categories (81\%), and 46/76 leaf categories (61\%) extracted from commits and developer-AI conversations, each mentioned by at least one interviewee. 

On top of that, they revealed how developers use AI agents, the benefits and challenges they perceive, and how AI-generated code issues affect the development process.

\textbf{Benefits.} All participants mentioned that AI coding tools considerably reduce manual coding effort, assist with rapid prototyping, increase overall productivity, and save time. This translates into more opportunities to focus on more important tasks, higher-level design, and big-picture thinking. 
At the same time, participants describe benefits that go beyond the automation of repetitive coding activities, such as boilerplate code generation. Three participants describe using AI for brainstorming ideas and architectures, understanding unfamiliar codebases and implementations from research papers, and selecting the most appropriate technology stacks. Another frequently mentioned benefit (5 out of 20 interviews) is that AI tools reduce barriers to learning new technologies and concepts, with one participant stressing that they now feel more confident when tackling novel tasks. Despite the many benefits of AI-assisted software development, there is a shared concern that these benefits can lead to overreliance on AI tools, resulting in a reduced understanding and ownership of the codebase, which inevitably creates serious risks.

\textbf{Challenges.} Nine participants mentioned the constant need to review, validate, and debug solutions generated by AI tools. They also note that this process sometimes consumes a significant portion of the time saved by automating development activities with AI. This is because AI tends to violate semantic properties and produce subtle bugs that are difficult to spot through simple inspection while still generating syntactically valid solutions (mentioned by 19 participants).

When it comes to the quality and maintainability of the produced code, some participants (3/20) feel that AI tools tend to prioritize quicker task completion over producing correct solutions. At the same time, there is an evident problem (3/20) that AI tools lose context and focus during long conversations, often forcing practitioners to restart from scratch. This inability to handle large contexts frequently leads to requested features being omitted from the produced implementation, a problem confirmed by the majority of participants (16/20). As a consequence, participants frequently adopt a strategy of breaking large tasks into smaller, reviewable steps and generating code incrementally instead of accepting large blocks of generated code at once. Several participants also shared frustration with AI becoming trapped in unproductive troubleshooting loops while remaining overconfident and ``boldly lying'', despite failing to solve the task. Almost everyone (16/20) acknowledged that they had at some point completely given up on a solution produced by AI tools because they failed to make it work.

Even when AI eventually produces a correct solution, the output may be unnecessarily complex, leading to technical debt accumulation. Some participants (3/20) working in specialized domains, such as hospitals, report that AI often struggles to produce correct output due to insufficient training data. They also note that AI tools are biased towards common patterns, technologies, and solutions, with one participant describing them as pattern matchers rather than reasoners.

Another difficulty inherent to the use of AI tools, shared by eight participants, is the significant effort required to learn how to communicate ideas and requirements effectively to AI tools. A related challenge, mentioned by five participants, is providing sufficient and consistent task context across different conversations with agents. This includes writing structured prompts that refer to specific requirements and repository conventions, and explicitly state constraints, preferred libraries, expected behavior, and edge cases. This indicates that prompt engineering is emerging as a critical developer skill. 

\textbf{Identification and handling of issues.} To identify problems in AI-generated code, all participants rely on a combination of quality assurance practices. Manual code review appears to be the dominant practice, complemented with functional, unit, and integration testing, followed by a validation of the code behavior in realistic environments. Many participants also report relying on linters, type checkers, pre-commit hooks, CI/CD pipelines, or existing test suites. Another frequently mentioned strategy is using AI itself as an independent reviewer, with participants asking another model to review and help diagnose or fix observed anomalies. To fix issues in AI-generated code, participants most commonly rely on an iterative approach, describing the observed problem, refining the requirements, and repeatedly asking the model to correct the implementation until a satisfactory solution is obtained. 

\textbf{Resemblance to human-made errors.} Finally, 8 out of 20 respondents believe that many of the faults in AI-generated code resemble those made by human developers, while almost everyone acknowledged the existence of AI-specific bugs. Three participants stated that AI-generated faults may resemble those observed in code produced by junior developers, but certainly not those of more experienced senior colleagues.


\subsection{Reasons Behind AI faults}
While it is critical to have a comprehensive overview of the different types of quality issues affecting AI-generated code, it is equally important to understand the reasons behind these faults. When inspecting developer-AI conversations as well as during our interviews, we collected the cases in which the reason behind a failure was clear, and we ended up with a list of twelve frequent causes that can be broadly divided into two general categories: LLM-induced reasons and prompt-induced ones. The former contains nine reasons that we manage to document for 116 of the quality issues in our taxonomy, and refer to cases in which the fault can be fully attributed on the LLM side. Examples of these reasons (full list in \cite{replication}) include: ``\emph{Gets lost in multi-turn conversation}'' (28 instances), ``\emph{Hallucinates APIs}'' (27), and ``\emph{Ignores provided instructions}'' (24). The latter (70 instances) comprises three fault triggers originating from a suboptimal user-provided prompts: ``\emph{Incomplete requirements specification}'' (43),  ``\emph{Presence of ambiguity}'' (24), and ``\emph{Wrong instructions}'' (3). All interview participants admitted that some code quality issues they encountered can be attributed to prompting mistakes.

\begin{table}[h]
	\centering
	\scriptsize
	\caption{Comparison with existing taxonomies.}
	\label{tab:comparison}
	\begin{tabular}{lr|rrr}
		\toprule
		\multirow{2}{*}{\textbf{Reference}} & \multicolumn{4}{c}{\textbf{Coverage level}}\\\cline{2-5}
		& \textbf{Overall} & \textbf{Root} & \textbf{Middle} & \textbf{Leaf}\\
		\midrule
		Liu \etal \cite{Liu:tosem2024} & 28\% & 50\% & 33\% & 24\%\\
		Wang \etal \cite{Wang:icse2025} & 19\% & 25\% & 26\% & 16\%\\
		Tambon \etal \cite{Tambon:emse2025} & 25\% & 38\% & 33\% & 22\%\\
		Dou \etal \cite{dou2025arxiv} & 30\% & 38\% & 33\% & 28\%\\
		Abbassi \etal \cite{Abbassi:icsme2025} & 27\% & 63\% & 41\% & 20\%\\
		\bottomrule
	\end{tabular}
\end{table}

\subsection{Comparison with Existing Taxonomies} \label{sec:comp-existing}
A distinctive feature of our taxonomy is that it was built (i) from quality issues documented in real software development practice, rather than from code produced by running LLMs on benchmarks; and (ii) by combining three complementary sources (\ie commits, developer-AI conversations, and interviews). In this section, we analyze how the issue types captured by our taxonomy differ from those in existing taxonomies \cite{Liu:tosem2024,Wang:icse2025,Tambon:emse2025,dou2025arxiv,Abbassi:icsme2025}. We exclude works do not directly classifying quality issues, but focusing on orthogonal aspects, such as the changes required to fix injected bugs \cite{Liu:tse2024}. Also, some prior taxonomies mix quality issues with their causes. For example, Liu \etal \cite{Liu:tosem2024} include both \emph{performance and efficiency} and \emph{task difficulty}; the former is a quality issue, while the latter is outside the scope of our taxonomy. We compare only categories representing quality issues.

Since the taxonomies differ in granularity, with ours substantially more detailed, we map at all levels: root, intermediate, and leaf nodes. Each quality issue type from prior work is mapped to the most specific matching node in our taxonomy. For example, `\emph{Compilation and Runtime Errors}'' \cite{Liu:tosem2024} maps to our root category `\emph{Build / compilation / static errors}'', whereas `\emph{Illegal Index}'' \cite{Liu:tosem2024} maps to our leaf node `\emph{Incorrect indexing / slicing}''. The complete mapping is available in our replication package \cite{replication}. Here, we summarize the main findings.

Our taxonomy covers all issue types documented in previous taxonomies. This is quite expected, as the largest existing taxonomy \cite{Liu:tosem2024} includes only 36 leaf nodes, compared to the \categories leaf nodes in ours (+144\%). \tabref{tab:comparison} reports, for each related work, the percentage of nodes in our taxonomy covered by that work, both overall and by node level (root, intermediate, and leaf). Existing taxonomies cover, on average, 25.7\% of our quality issue types (min: 19\%, max: 30\%). Even when considering the union of all issue types reported in prior work, our taxonomy  includes 50 AI-generated quality issue types previously undocumented (\eg \emph{Hard-coded invalid value/placeholder}).

%% file: threats.tex
\section{Threats to Validity} \label{sec:threats}

\textbf{Construct validity.} The commit analysis is limited to documented fixes of AI-introduced issues, and therefore misses silent fixes. Still, this methodology is commonly adopted in studies of technical debt \cite{PotdarS14} and generative AI usage \cite{tufano:msr2024,PepeNMCBP26,10.1145/3771937}. Another threat concerns the identification of AI errors in commits and conversations, which was performed by the authors rather than the original developers; we mitigated it through multiple annotators. 

To maximize consistency in the labeling of artifacts, after each inspection round (\tabref{tab:new_categories_per_round}), the authors reviewed and consolidated the labels and previous assignments. Whenever a new label was introduced, we assessed whether it overlapped with, refined, or partially subsumed an existing label. If so, we revisited the instances previously assigned to the related existing label(s), together with the instances motivating the new label, and reconsidered their coding. 

Finally, interview data may reflect participants' perceptions rather than objective facts.

\textbf{Internal validity.} For the mining study, threats may stem from: (i) the keywords used as filters, which may not be exhaustive despite being carefully selected (\secref{sec:design}); (ii) the non-random selection of ChatGPT conversation files, limited to the first 1,000 retrieved; and (iii) the exclusion of conversations being longer than 19 rounds and/or without explicit AI apologies. To study the impact of the conversation-selection choices, we randomly sampled 30 conversations longer than 19 rounds that contain an apology signal, thus isolating the effect of the length threshold, and 30 conversations within the retained length range but without any apology signal, thus isolating the effect of the apology-based filter. Each conversation was independently inspected by two authors using the same coding procedure adopted in the main study, including conflict resolution when needed. The 30 long conversations with apology signal featured a total of 888 rounds: 21 conversations contained at least one relevant quality issue, yielding 128 issue instances, all mapped to existing taxonomy categories. Among the 30 conversations without apology signals, two contained relevant issues, again all mapped in our taxonomy. Overall, no new quality-issue types emerged. These results suggest that the current taxonomy is reasonably robust to these two conversation-selection choices. 

Another threat concerns subjectivity in the coding tasks. We mitigated it through two independent annotators per artifact, an additional pair of annotators for conflict resolution, and multiple taxonomy-refinement stages involving all authors. For the commit analysis, when both evaluators assigned at least one label, they selected exactly the same set of labels in 40\% of cases and at least one identical leaf-level label in 44\% of cases; when considering the hierarchical structure of the taxonomy, their coding agreed at least at the root-category level in 65\% of cases. For the conversation analysis, these numbers are 42\%, 47\%, and 76\%, respectively. Finally, for the interviews, we got 40\%, 57\%, and 81\%. These figures concern the initial independent coding, before reconciliation, and should be interpreted in the context of an inductive coding process in which the taxonomy itself was progressively refined. We also examined all disagreement-resolution outcomes. For commits, the final coding was identical to that proposed by one (or both) evaluator in 84\% of cases, used a combination of all proposed labels in 4\%, and introduced at least one new/refined label in 12\%. For conversations, the corresponding numbers were 83\%, 3\%, and 14\%. For interviews, 75\%, 10\%, and 15\%.

For the interview study, a further threat is self-selection bias: participants may have been especially interested in the topic. However, since our goal was to collect experiences of issues encountered when using generative AI for coding, we consider this threat limited.

\textbf{External validity.} While we mined commits mentioning a broad set of AI tools, our search does not cover all available technologies. The conversation analysis is limited to ChatGPT and may not generalize to other tools. Finally, the interview findings are bounded by the set of participants.

%% file: conclusion.tex
\section{Conclusion \& Future Work} \label{sec:conclusion}
We presented a taxonomy of quality issues in AI-generated code derived from three complementary sources: fixing commits, developer-AI conversations, and practitioner interviews. The taxonomy comprises \categories types of quality issues, making it, to the best of our knowledge, the most comprehensive taxonomy on this topic to date. Developers can use the taxonomy to inform code review and testing practices for AI-generated code. Researchers can use it as a foundation for designing more representative evaluation benchmarks and quality assurance techniques for AI coding assistants, which we plan to investigate in future work.

%% file: acknowledgements.tex
\section{Acknowledgements} \label{sec:acknow}
Bavota and Tufano acknowledge the financial support of the Swiss National Science Foundation for the PARSED project (SNF Project No. 219294). 
Tonella and Humbatova acknowledge the financial support of the Swiss National Science Foundation for the Toposcope project (SNF Project No. 214989). 
Jahangirova acknowledges the financial support of the ITEA grant GENIUS (Project No. 23026).
Bifolco acknowledges the financial support of the PNRR DM 118/2023 Italian Grant for Ph.D. scholarships.
We thank all the interview participants for their time and valuable insights.

%% file: main.bib
@misc{Streamlit,
  author  = {{Streamlit}},
  title   = {Streamlit: A Faster Way to Build and Share Data Apps},
  year    = {2026},
  url     = {https://streamlit.io/},
  urldate = {2026-06-26}
}

@article{Etikan2016ConvenienceSampling,
  title={Comparison of convenience sampling and purposive sampling},
  author={Etikan, Ilker and Musa, Sulaiman Abubakar and Alkassim, Rukayya Sunusi and others},
  journal={American journal of theoretical and applied statistics},
  volume={5},
  number={1},
  pages={1--4},
  year={2016},
  publisher={New York}
}

@article{PepeNMCBP26,
	author       = {Federica Pepe and
	Vittoria Nardone and
	Antonio Mastropaolo and
	Gerardo Canfora and
	Gabriele Bavota and
	Massimiliano {Di Penta}},
	title        = {Datasets, bias, licenses, and terms of use: {A} large and longitudinal
	study on the documentation of hugging face machine learning models},
	journal      = {Empir. Softw. Eng.},
	volume       = {31},
	number       = {4},
	pages        = {95},
	year         = {2026},
	doi          = {10.1007/S10664-026-10843-1},
	bibsource    = {dblp computer science bibliography, https://dblp.org}
}

@inproceedings{PotdarS14,
	author={Potdar, Aniket and Shihab, Emad},
  booktitle={2014 IEEE International Conference on Software Maintenance and Evolution}, 
  title={An Exploratory Study on Self-Admitted Technical Debt}, 
  year={2014},
  pages={91-100},
  doi={10.1109/ICSME.2014.31}
}

@article{10.1145/3771937,
	author = {Kashif, Syed Mohammad and Liang, Peng and Tahir, Amjed},
	title = {On Developers’ Self-Declaration of AI-Generated Code: An Analysis of Practices},
	year = {2025},
	publisher = {Association for Computing Machinery},
	address = {New York, NY, USA},
	issn = {1049-331X},
	url = {https://doi.org/10.1145/3771937},
	doi = {10.1145/3771937},
	note = {Just Accepted},
	journal = {ACM Trans. Softw. Eng. Methodol.},
	month = oct
}

@article{braun2006using,
	title={Using thematic analysis in psychology},
	author={Braun, Virginia and Clarke, Victoria},
	journal={Qualitative Research in Psychology},
	volume={3},
	number={2},
	pages={77--101},
	year={2006},
	publisher={Taylor \& Francis},
	doi={10.1191/1478088706qp063oa}
}

@Article{kappa,
	title={A coefficient of agreement for nominal scales},
  author={Cohen, Jacob},
  journal={Educational and psychological measurement},
  volume={20},
  number={1},
  pages={37--46},
  year={1960},
  publisher={Sage Publications Sage CA: Thousand Oaks, CA}
}

@misc{https://github.com/Stensel8/Scripts/commit/1e270a21805997353c9ca038a89770e9a0f8cd46,
  year = {2025},
  title = {Update nginx\_installer.sh to use OpenSSL version 3.5.2 (\#30) · Stensel8/Scripts@1e270a2 · GitHub},
  url		= {https://github.com/Stensel8/Scripts/commit/1e270a21805997353c9ca038a89770e9a0f8cd46},
  urldate = {2026-06-05}
}

@misc{https://chat.openai.com/share/b78b8a63-fb6e-41c2-a874-3c682b61578f,
  year = {2026},
  title = {ChatGPT - OCR Image Text Recognition},
  url		= {https://chat.openai.com/share/b78b8a63-fb6e-41c2-a874-3c682b61578f},
  urldate = {2026-06-05}
}

@misc{https://github.com/Tonypsilon/rankify-api/commit/541c6962606f21a13ed2d74686e0d64619501901,
  year = {2025},
  title = {Fix ordering issue introduced by Copilot: Was sorting poll options alphabetically for no reason. · Tonypsilon/rankify-api@541c696 · GitHub},
  url		= {https://github.com/Tonypsilon/rankify-api/commit/541c6962606f21a13ed2d74686e0d64619501901},
  urldate = {2026-06-05}
}

@misc{https://chatgpt.com/share/68121cb1-e234-8005-a86b-5f924a0f7784,
  year = {2026},
  title = {ChatGPT - Stress symbol recovery algorithm},
  url		= {https://chatgpt.com/share/68121cb1-e234-8005-a86b-5f924a0f7784},
  urldate = {2026-06-05}
}

@misc{https://chat.openai.com/share/6c356c82-3607-4c92-8271-99144239b1ae,
  year = {2026},
  title = {ChatGPT - Automate Plugin Data Extraction},
  url		= {https://chat.openai.com/share/6c356c82-3607-4c92-8271-99144239b1ae},
  urldate = {2026-06-05}
}

@misc{https://github.com/microsoft/TypeAgent/commit/ff32d75e4cbd80726549a2f0bfaaa9b4ea677886,
  year = {2025},
  title = {[Spelunker] Fix bugs introduced by Copilot (and my poor review skills) (\#698) · microsoft/TypeAgent@ff32d75 · GitHub},
  url		= {https://github.com/microsoft/TypeAgent/commit/ff32d75e4cbd80726549a2f0bfaaa9b4ea677886},
  urldate = {2026-06-05}
}

@misc{https://chatgpt.com/share/6713d47d-4960-8003-9869-293729ea5b8c,
  year = {2026},
  title = {ChatGPT - NohBoard Caps Lock Issue},
  url		= {https://chatgpt.com/share/6713d47d-4960-8003-9869-293729ea5b8c},
  urldate = {2026-06-05}
}

@misc{https://chat.openai.com/share/821d8f25-f37c-4cfc-8ca5-0ff7200e3f04,
  year = {2026},
  title = {ChatGPT - F\# Computation Expression},
  url		= {https://chat.openai.com/share/821d8f25-f37c-4cfc-8ca5-0ff7200e3f04},
  urldate = {2026-06-05}
}

@misc{https://github.com/leonzdev/tether-router-monitor/commit/9e4b69e695fe2c40e8794b1a9cab4a18b324163a,
  year = {2023},
  title = {Update main.go · leonzdev/tether-router-monitor@9e4b69e · GitHub},
  url		= {https://github.com/leonzdev/tether-router-monitor/commit/9e4b69e695fe2c40e8794b1a9cab4a18b324163a},
  urldate = {2026-06-05}
}

@misc{https://chatgpt.com/share/d85c4695-4b71-4e13-a4b2-51f14031198f,
  year = {2026},
  title = {ChatGPT - Override 'isSource' for 'a'},
  url		= {https://chatgpt.com/share/d85c4695-4b71-4e13-a4b2-51f14031198f},
  urldate = {2026-06-05}
}

@misc{https://github.com/Ztgunderson/ECE455/commit/211808c1b4c385f42de4b17f1de791e8f060a150,
  year = {2026},
  title = {fixing syntax issues from using gemini · Ztgunderson/ECE455@211808c · GitHub},
  url		= {https://github.com/Ztgunderson/ECE455/commit/211808c1b4c385f42de4b17f1de791e8f060a150},
  urldate = {2026-06-05}
}

@misc{https://chatgpt.com/share/686a0646-7e1c-8002-a487-154858aa3157,
  year = {2026},
  title = {ChatGPT - Long Parameter List \#1 - AudioReference},
  url		= {https://chatgpt.com/share/686a0646-7e1c-8002-a487-154858aa3157},
  urldate = {2026-06-05}
}

@misc{https://chatgpt.com/share/686e2bdc-f6c4-8008-9e00-06378c2bdbc1,
  year = {2026},
  title = {ChatGPT - 10. Long Parameter List Refactoring},
  url		= {https://chatgpt.com/share/686e2bdc-f6c4-8008-9e00-06378c2bdbc1},
  urldate = {2026-06-05}
}

@misc{https://github.com/s3bc40/moccasin/commit/461ad6998a9d4ba0b94a27ac3684e356edeb04c5,
  year = {2025},
  title = {fix: error from copilot duplicate code · s3bc40/moccasin@461ad69 · GitHub},
  url		= {https://github.com/s3bc40/moccasin/commit/461ad6998a9d4ba0b94a27ac3684e356edeb04c5},
  urldate = {2026-06-05}
}

@misc{https://github.com/podaac/bignbit/commit/f68fceb763ae7c24fa6f7439211fe739c580577e,
  year = {2025},
  title = {fix bug introduced by Copilot · podaac/bignbit@f68fceb · GitHub},
  url		= {https://github.com/podaac/bignbit/commit/f68fceb763ae7c24fa6f7439211fe739c580577e},
  urldate = {2026-06-05}
}

@misc{https://chat.openai.com/share/e9f4664e-50a5-40c1-8604-befe89a2dd36,
  year = {2026},
  title = {ChatGPT - Add Task Event Handler},
  url		= {https://chat.openai.com/share/e9f4664e-50a5-40c1-8604-befe89a2dd36},
  urldate = {2026-06-05}
}

@misc{https://chatgpt.com/share/c4ca8622-9b01-4189-9e29-7414ddf79238,
  year = {2026},
  title = {ChatGPT - Go Scripting with goscript},
  url		= {https://chatgpt.com/share/c4ca8622-9b01-4189-9e29-7414ddf79238},
  urldate = {2026-06-05}
}

@misc{https://chat.openai.com/share/608ed300-6714-4c74-be6e-022e80b05207,
  year = {2026},
  title = {ChatGPT - Fix Tailwind CSS Config},
  url		= {https://chat.openai.com/share/608ed300-6714-4c74-be6e-022e80b05207},
  urldate = {2026-06-05}
}

@misc{https://chat.openai.com/share/b0dabec7-a903-4d50-b300-89801ec8e31d,
  year = {2026},
  title = {ChatGPT - Complex Node.js Function \& Tests},
  url		= {https://chat.openai.com/share/b0dabec7-a903-4d50-b300-89801ec8e31d},
  urldate = {2026-06-05}
}

@article{Seaman:1999,
 author = {Seaman, Carolyn B.},
 title = {Qualitative Methods in Empirical Studies of Software Engineering},
 journal = {IEEE Trans. Softw. Eng.},
 issue_date = {July 1999},
 volume = {25},
 number = {4},
 month = jul,
 year = {1999},
 issn = {0098-5589},
 pages = {557--572},
 numpages = {16},
 doi = {10.1109/32.799955},
 acmid = {322687},
 publisher = {IEEE Press},
 address = {Piscataway, NJ, USA},
}

@url{GHArchive,
	title = {\url{https://www.gharchive.org/}},
	year = {2021},
}

@url{GitHubSearch,
	title = {\url{https://docs.github.com/en/rest/search/search}},
	year = {2024},
}

@inproceedings{Xiao:MSR2024,
  author = {Xiao, Tao and Treude, Christoph and Hata, Hideaki and Matsumoto, Kenichi},
  title = {DevGPT: Studying Developer-ChatGPT Conversations},
  year = {2024},
  isbn = {9798400705878},
  publisher = {Association for Computing Machinery},
  address = {New York, NY, USA},
  doi = {10.1145/3643991.3648400},
  booktitle = {Proceedings of the 21st International Conference on Mining Software Repositories},
  pages = {227–230},
  numpages = {4},
  location = {Lisbon, Portugal},
  series = {MSR '24}
}

@INPROCEEDINGS {Jesse:msr2023,
  author={Jesse, Kevin and Ahmed, Toufique and Devanbu, Premkumar T. and Morgan, Emily},
  booktitle={2023 IEEE/ACM 20th International Conference on Mining Software Repositories (MSR)}, 
  title={Large Language Models and Simple, Stupid Bugs}, 
  year={2023},
  pages={563-575},
  doi={10.1109/MSR59073.2023.00082}
}

@INPROCEEDINGS {Wang:icse2025,
  author={Wang, Zhijie and Zhou, Zijie and Song, Da and Huang, Yuheng and Chen, Shengmai and Ma, Lei and Zhang, Tianyi},
  booktitle={2025 IEEE/ACM 47th International Conference on Software Engineering (ICSE)}, 
  title={Towards Understanding the Characteristics of Code Generation Errors Made by Large Language Models}, 
  year={2025},
  volume={},
  number={},
  pages={2587-2599},
  doi={10.1109/ICSE55347.2025.00180}
}

@ARTICLE{Liu:tse2024,
  author={Liu, Zhijie and Tang, Yutian and Luo, Xiapu and Zhou, Yuming and Zhang, Liang Feng},
  journal={IEEE Transactions on Software Engineering}, 
  title={No Need to Lift a Finger Anymore? Assessing the Quality of Code Generation by ChatGPT}, 
  year={2024},
  volume={50},
  number={6},
  pages={1548-1584}}

@inproceedings{Moratis:msr2024,
  author = {Moratis, Konstantinos and Diamantopoulos, Themistoklis and Nastos, Dimitrios-Nikitas and Symeonidis, Andreas},
  title = {Write me this Code: An Analysis of ChatGPT Quality for Producing Source Code},
  year = {2024},
  isbn = {9798400705878},
  publisher = {Association for Computing Machinery},
  address = {New York, NY, USA},
  doi = {10.1145/3643991.3645070},
  booktitle = {Proceedings of the 21st International Conference on Mining Software Repositories},
  pages = {147–151},
  numpages = {5},
  location = {Lisbon, Portugal},
  series = {MSR '24}
}

@inproceedings{Siddiq:msr2024,
  author       = {Mohammed Latif Siddiq and
                  Lindsay Roney and
                  Jiahao Zhang and
                  Joanna C. S. Santos},
  title        = {Quality Assessment of ChatGPT Generated Code and their Use by Developers},
  booktitle    = {21st {IEEE/ACM} International Conference on Mining Software Repositories,
                  {MSR} 2024},
  pages        = {152--156},
  publisher    = {{ACM}},
  year         = {2024},
  doi = {10.1145/3643991.3645071},

}

@inproceedings{tufano:msr2024,
  author       = {Rosalia Tufano and
                  Antonio Mastropaolo and
                  Federica Pepe and
                  Ozren Dabic and
                  Massimiliano Di Penta and
                  Gabriele Bavota},
  title        = {Unveiling ChatGPT's Usage in Open Source Projects: {A} Mining-based
                  Study},
  booktitle    = {21st {IEEE/ACM} International Conference on Mining Software Repositories,
                  {MSR} 2024},
  pages        = {571--583},
  publisher    = {{ACM}},
  year         = {2024},
  doi          = {10.1145/3643991.3644918},
}

@inproceedings{Fadi:ceur,
  author       = {Fadi Barb{\`{a}}ra and
                  Emanuele Antonio Napoli and
                  Valentina Gatteschi and
                  Claudio Schifanella},
  editor       = {Massimo Bartoletti and
                  Claudio Schifanella and
                  Andrea Vitaletti},
  title        = {Automatic Smart Contract Generation Through LLMs: When The Stochastic
                  Parrot Fails},
  booktitle    = {Proceedings of the Sixth Distributed Ledger Technology Workshop {(DLT}
                  2024)},
  series       = {{CEUR} Workshop Proceedings},
  year         = {2024}
}

@article{Cui2024Productivity,
	author = {Cui, Kevin Zheyuan and Demirer, Mert and Jaffe, Sonia and Musolff, Leon and Peng, Sida and Salz, Tobias},
	journal = {An MIT Exploration of Generative AI},
	year = {2024},
	month = {mar 27},
	note = {https://mit-genai.pubpub.org/pub/v5iixksv},
	publisher = {MIT},
	title = {The {Productivity} {Effects} of {Generative} {AI}: Evidence from a {Field} {Experiment} with {GitHub} {Copilot}},
  doi = {10.21428/e4baedd9.3ad85f1c},
}

@inproceedings{CoderEval,
  author = {Yu, Hao and Shen, Bo and Ran, Dezhi and Zhang, Jiaxin and Zhang, Qi and Ma, Yuchi and Liang, Guangtai and Li, Ying and Wang, Qianxiang and Xie, Tao},
  title = {CoderEval: A Benchmark of Pragmatic Code Generation with Generative Pre-trained Models},
  year = {2024},
  isbn = {9798400702174},
  publisher = {Association for Computing Machinery},
  address = {New York, NY, USA},
  doi = {10.1145/3597503.3623316},
  booktitle = {Proceedings of the IEEE/ACM 46th International Conference on Software Engineering},
  articleno = {37},
  numpages = {12},
  location = {Lisbon, Portugal},
  series = {ICSE '24}
}

@article{Tambon:emse2025,
  title={Bugs in large language models generated code: An empirical study},
  author={Tambon, Florian and Moradi-Dakhel, Arghavan and Nikanjam, Amin and Khomh, Foutse and Desmarais, Michel C and Antoniol, Giuliano},
  journal={Empirical Software Engineering},
  volume={30},
  number={3},
  pages={65},
  year={2025},
  publisher={Springer},
  doi={10.1007/s10664-025-10614-4}
}

@article{Liu:tosem2024,
  author = {Liu, Yue and Le-Cong, Thanh and Widyasari, Ratnadira and Tantithamthavorn, Chakkrit and Li, Li and Le, Xuan-Bach D. and Lo, David},
  title = {Refining ChatGPT-Generated Code: Characterizing and Mitigating Code Quality Issues},
  year = {2024},
  publisher = {Association for Computing Machinery},
  address = {New York, NY, USA},
  journal = {ACM Trans. Softw. Eng. Methodol.},
  volume = {33},
  number = {5},
  month = jun,
  articleno = {116},
  numpages = {26},
  doi = {10.1145/3643674},
}

@article{Fu:tosem2025,
  author = {Fu, Yujia and Liang, Peng and Tahir, Amjed and Li, Zengyang and Shahin, Mojtaba and Yu, Jiaxin and Chen, Jinfu},
  title = {Security Weaknesses of Copilot-Generated Code in GitHub Projects: An Empirical Study},
  year = {2025},
  issue_date = {November 2025},
  publisher = {Association for Computing Machinery},
  address = {New York, NY, USA},
  volume = {34},
  number = {8},
  issn = {1049-331X},
  doi = {10.1145/3716848},
  journal = {ACM Trans. Softw. Eng. Methodol.},
  articleno = {218},
  numpages = {34},
}

@inproceedings{Mousavi:ccs2024,
  author = {Mousavi, Zahra and Islam, Chadni and Moore, Kristen and Abuadbba, Alsharif and Babar, M. Ali},
  title = {An Investigation into Misuse of Java Security APIs by Large Language Models},
  year = {2024},
  booktitle = {Proceedings of the 19th ACM Asia Conference on Computer and Communications Security},
  pages = {1299–1315},
  numpages = {17},
  series = {ASIA CCS '24},
  doi = {10.1145/3634737.3661134},
}

@article{Zhang:issta2025,
  author = {Zhang, Ziyao and Wang, Chong and Wang, Yanlin and Shi, Ensheng and Ma, Yuchi and Zhong, Wanjun and Chen, Jiachi and Mao, Mingzhi and Zheng, Zibin},
  title = {LLM Hallucinations in Practical Code Generation: Phenomena, Mechanism, and Mitigation},
  year = {2025},
  issue_date = {July 2025},
  publisher = {Association for Computing Machinery},
  address = {New York, NY, USA},
  volume = {2},
  number = {ISSTA},
  doi = {10.1145/3728894},
  journal = {Proc. ACM Softw. Eng.},
  articleno = {ISSTA022},
  numpages = {23},
}

@article{dou2025arxiv,
  title={What is wrong with your code generated by large language models? An extensive study},
  author={Dou, Shihan and Jia, Haoxiang and Wu, Shenxi and Zheng, Huiyuan and Wu, Muling and Tao, Yunbo and Zhang, Ming and Chai, Mingxu and Fan, Jessica and Xi, Zhiheng and others},
  journal={Science China Information Sciences},
  volume={69},
  number={1},
  pages={112107},
  year={2026},
  publisher={Springer}
}

@misc{liu:arxiv2026,
      title={Beyond Functional Correctness: Exploring Hallucinations in LLM-Generated Code}, 
      author={Fang Liu and Yang Liu and Lin Shi and Zhen Yang and Li Zhang and Xiaoli Lian and Zhongqi Li and Yuchi Ma},
      year={2026},
      eprint={2404.00971},
      archivePrefix={arXiv},
      primaryClass={cs.SE},
      url={https://arxiv.org/abs/2404.00971}, 
}

@misc{gao2025arxiv,
      title={A Survey of Bugs in AI-Generated Code}, 
      author={Ruofan Gao and Amjed Tahir and Peng Liang and Teo Susnjak and Foutse Khomh},
      year={2025},
      eprint={2512.05239},
      archivePrefix={arXiv},
      primaryClass={cs.SE},
      url={https://arxiv.org/abs/2512.05239}, 
}

@inproceedings{Abbassi:icsme2025,
  author={Abbassi, Altaf Allah and Da Silva, Leuson and Nikanjam, Amin and Khomh, Foutse},
  booktitle={2025 IEEE International Conference on Software Maintenance and Evolution (ICSME)}, 
  title={A Taxonomy of Inefficiencies in LLM-Generated Python Code}, 
  year={2025},
  pages={393-404},
  doi={10.1109/ICSME64153.2025.00043}
}

@inproceedings{cotroneo:icpc2024,
  author       = {Domenico Cotroneo and
                  Cristina Improta and
                  Pietro Liguori and
                  Roberto Natella},
  title        = {Vulnerabilities in {AI} Code Generators: Exploring Targeted Data Poisoning
                  Attacks},
  booktitle    = {Proceedings of the 32nd {IEEE/ACM} International Conference on Program
                  Comprehension, {ICPC}},
  pages        = {280--292},
  publisher    = {{ACM}},
  year         = {2024},
  doi          = {10.1145/3643916.3644416},
}

@misc{amazon-meeting,
  title  = {Amazon holds engineering meeting following AI-related outages},
  howpublished = {\url{https://www.ft.com/content/7cab4ec7-4712-4137-b602-119a44f771de}},
  author       = {Financial Times},
  year         = {2026},
  note = {{Last Accessed: June 2026}},
}

@misc{chatgpt,
	Title = {ChatGPT},
  howpublished = {\url{https://openai.com/blog/chatgpt}},
}

@misc{copilot,
  title  = {{GitHub Copilot -- Your AI pair programmer}},
  howpublished = {\url{https://github.com/features/copilot/}},
  note         = {Accessed: 2024-03-10}
}

@misc{replication,
  title  = {Replication Package},
  howpublished = {\url{https://github.com/RosaliaTufano/AIgenerated_code_issues}},
  year = {[n.d.]}
}

@article{humaneval,
  title={Evaluating large language models trained on code},
  author={Chen, Mark and Tworek, Jerry and Jun, Heewoo and Yuan, Qiming and Pinto, Henrique Ponde de Oliveira and Kaplan, Jared and Edwards, Harri and Burda, Yuri and Joseph, Nicholas and Brockman, Greg and others},
  journal={arXiv preprint arXiv:2107.03374},
  year={2021}
}

@article{evalplus,
  author = {Liu, Jiawei and Xia, Chunqiu Steven and Wang, Yuyao and ZHANG, LINGMING},
  booktitle = {Advances in Neural Information Processing Systems},
  editor = {A. Oh and T. Naumann and A. Globerson and K. Saenko and M. Hardt and S. Levine},
  pages = {21558--21572},
  publisher = {Curran Associates, Inc.},
  title = {Is Your Code Generated by ChatGPT Really Correct? Rigorous Evaluation of Large Language Models for Code Generation},
  url = {https://proceedings.neurips.cc/paper_files/paper/2023/file/43e9d647ccd3e4b7b5baab53f0368686-Paper-Conference.pdf},
  volume = {36},
  year = {2023}
}

@inproceedings{robbes2026promises,
  title={Promises, Perils, and (Timely) Heuristics for Mining Coding Agent Activity},
  author={Robbes, Romain and Matricon, Th{\'e}o and Degueule, Thomas and Hora, Andre and Zacchiroli, Stefano},
  booktitle={MSR 2026-23rd IEEE Working Conference on Mining Software Repositories},
  year={2026}
}
